\documentclass{article}
\usepackage{amssymb}
\usepackage{bm}
\usepackage[dvips]{graphicx}
\begin{document}

\title{Current-driven spin injection\\ from a probe to a ferromagnetic film}

\author{S. G. Chigarev, E. M. Epshtein\thanks{E-mail: eme253@ms.ire.rssi.ru},
Yu. V. Gulyaev,\\ A. I. Panas, V. G. Shofman, P. E.
Zilberman\\ \\
V.A. Kotelnikov Institute of Radio Engineering and Electronics\\
of the Russian Academy of Sciences, 141190 Fryazino, Russia}

\date{}

\maketitle

\abstract{The distribution is calculated of the electron spin polarization
under current-driven spin injection from a probe to a ferromagnetic film. It is shown
that the main parameters determining difference of the spin polarization
from the equilibrium value are the current density and the spin
polarization of the probe material, while the relation between the probe
diameter and the spin diffusion length influences the result very weakly, to a certain extent.
A possibility is shown of reaching inverse population of the spin subbands
at distances from the probe boundary comparable with the spin diffusion
length.}

\section{Introduction}\label{section1}
The current-induced spin injection, i.e., appearing nonequilibrium spin polarization near
boundary between two conductors under current flowing
through~\cite{Aronov}, is one of the main spintronic effects together with
the effect of spin torque transfer from conduction electrons to
lattice~\cite{Slonczewski,Berger}. The spin injection manifests itself as
breaking the thermal equilibrium between spin subbands.

Nowadays, the spin torque transfer effect has been studied with much more
details than the spin injection. However, an attractive problem occurs
that relates with reaching high injection level. We mean a possibility of
creating inverse population of the spin subbands in a ferromagnet with
laser effect in THz and IR ranges ($10^{12}$--$10^{14}$
Hz)~\cite{Kadigrobov1}--\cite{Gulyaev2}. The main obstacle for this
idea is necessity of high current density $j\ge10^9$ A/cm$^2$.

A scheme was proposed for reaching high current density in a microprobe--thin film
system~\cite{Chigarev}. If the film thickness $h$ is small compared to the
probe radius $R$, then the current density in the film near the probe is
$R/2h$ times the current density in the film. So a problem appears of
calculating spin injection in such a system as a function of the system
parameters and the current value. Such a problem is considered in the
present work.

\section{The model and main equations}\label{section2}
Let us consider a ferromagnetic film of $h$ thickness which current $I$ is
carried to by means of a cylindrical probe of $R$ radius. The current in
the film is spin-polarized, so that its flowing is accompanied with
appearing a nonequilibrium spin polarization near the probe at distances
comparable with the spin diffusion length $l$.

The spin polarization is defined as
\begin{equation}\label{1}
  P=\frac{n_+-n_-}{n},
\end{equation}
where $n_+$ and $n_-$ are the partial densities of the conduction
electrons with spin magnetic moments aligned parallel and antiparallel to
the crystal lattice magnetization vector, respectively, $n=n_++n_-$ is the total
density of the conduction density, which is assumed to be constant. It
follows from Eq.~(\ref{1})
\begin{equation}\label{2}
  n_\pm=\frac{n}{2}(1\pm P).
\end{equation}

The spin polarization distribution under steady conditions is determined
by the spin current continuity equation
\begin{equation}\label{3}
  \nabla\mathbf J=-\frac{\hbar n}{2}\frac{P-\bar P}{\tau},
\end{equation}
where $\bar P$ is the equilibrium spin polarization, $\tau$ is the
relaxation time of the longitudinal (collinear with the lattice
magnetization) component of the electron spin polarization,
\begin{equation}\label{4}
  \mathbf J=\frac{\hbar}{2e}(\mathbf j_+-\mathbf j_-)
\end{equation}
is the spin current density
\begin{equation}\label{5}
  \mathbf j_\pm=e\mu_\pm n_\pm\mathbf E-eD_\pm\nabla n_\pm,
\end{equation}
are the partial current densities created by the electrons with two
opposite spin directions, $\mathbf E$ is the electric field, $\mu_\pm$ and
$D_\pm$ are the partial mobilities and the partial diffusion constants,
respectively.

The electric field $\mathbf E$ can be expressed via the total current
density $\mathbf j=\mathbf j_++\mathbf j_-$. The spin current density~(\ref{4}) takes the form
\begin{equation}\label{6}
  \mathbf J=\frac{\hbar}{2e}\{Q(P)\mathbf j-enD(P)\nabla P\},
\end{equation}
where
\begin{equation}\label{7}
  Q(P)=\frac{\mu_+-\mu_-+(\mu_++\mu_-)P}{\mu_++\mu_-+(\mu_+-\mu_-)P},
\end{equation}
\begin{equation}\label{8}
  D(P)=\frac{\mu_+D_-+\mu_-D_++(\mu_+D_--\mu_-D_+)P}{\mu_++\mu_-+(\mu_+-\mu_-)P}.
\end{equation}
Substitution of Eq.~(\ref{6}) to Eq.~(\ref{3}) gives an equation for the
spin polarization $P$ which is substantially nonlinear as follows from
Eqs.~(\ref{2}),~(\ref{7}) and~(\ref{8}). Besides the explicit
linear-fractional dependence of $Q(P)$ and $D(P)$ coefficients on $P$, the partial
mobilities and diffusion constants of a degenerate electron gas (metal) depend
in general on the Fermi quasilevels of the spin subbands and,
subsequently, on the partial densities $n_\pm$, which, in their turn, are
expressed via spin polarization $P$ (see Eq.~(\ref{2})). The form of that
dependence is determined by many factors, such as the form of the spin
subbands (dispersion law), the carrier scattering mechanism, etc.

In many cases, such as the problem of the switching magnetic configuration
by spin-polarized current, a linear approximation in the current density
$\mathbf j$ and proportional to it nonequilibrium spin polarization $\Delta P=P-\bar P$ appears to
be sufficient. In this approximation, the spin polarization $P$ in $Q(P)$ and $D(P)$
coefficients is replaced with its equilibrium value $\bar P$, so that the
coefficients mentioned take constant values $\bar Q\equiv Q(\bar P)$ and $\bar D\equiv D(\bar
P)$, and Eq.~(\ref{3}) becomes linear equation, namely,
\begin{equation}\label{9}
    \nabla^2P-\frac{P-\bar P}{\bar l^2}=0,
\end{equation}
where $\bar l=\sqrt{\bar D\tau}$.

The linear approximation becomes invalid under high spin injection
corresponding to the spin subband inverse population, when $P<0$, so that
$\Delta P<0$, $|\Delta P|>\bar P$~\cite{Gulyaev1}. However, the situation is
simplified noticeably if it is supposed that the
carriers in both spin subbands have the same mobilities and diffusion constants,
$\mu_-=\mu_+\equiv\mu$, $D_-=D_+\equiv D$. In such a case, we have
$Q(P)=P$, $D(P)=D$. The substitution of Eq.~(\ref{6}) into Eq.~(\ref{3}) with
the electric charge conservation condition $\nabla\mathbf j=0$ taking into
account gives the equation
\begin{equation}\label{10}
  \nabla^2P-\frac{\mathbf j\nabla P}{j_Dl}-\frac{P-\bar P}{l^2}=0,
\end{equation}
where
\begin{equation}\label{11}
  l=\sqrt{D\tau},\quad j_D=enD/l=enl/\tau.
\end{equation}
The form of this equation depends on
neither the carrier degeneration, nor the dispersion law, nor the
scattering mechanism. The solution of such simplified problem, without
having any pretension to obtaining quantitative results for particular
materials, allows to find a qualitative picture and estimate the orders of
magnitude.

\section{Spin polarization distribution}\label{section3}
Under spin injection conditions, the spin polarization differs from its
equilibrium value at the distances from the injector comparable with the
spin diffusion length $l$. If the film lateral size is large in comparison
with that length, the current density distribution may be considered as
axially symmetrical one irrespective of the geometry of the other
electrode closing the electric circuit. It follows from the electric
charge conservation condition
\begin{equation}\label{12}
  \nabla\mathbf j=\frac{1}{r}\frac{d}{dr}(rj)=0
\end{equation}
that the current density distribution in the film near the probe takes the
form
\begin{equation}\label{13}
  j(r)=j(R)\frac{R}{r}=\frac{I}{2\pi hr},
\end{equation}
where $r$ is the distance from the probe axis, the other notations being
indicated above.

The substitution of Eq.~(\ref{13}) into Eq.~(\ref{10}) gives the following
equation in polar coordinates:
\begin{equation}\label{14}
  \frac{d^2P}{dr^2}+(1-2\nu)\frac{1}{r}\frac{dP}{dr}-\frac{P-\bar P}{l^2}=0,
\end{equation}
where
\begin{equation}\label{15}
  \nu=\frac{1}{2}\frac{R}{l}\frac{j(R)}{j_D},
\end{equation}
The complete solution of Eq.~(\ref{14}) has the form~\cite{Kamke}
\begin{equation}\label{16}
  P(r)=\bar P+\left(\frac{r}{l}\right)^\nu\left[C_1I_\nu\left(\frac{r}{l}\right)
  +C_2K_\nu\left(\frac{r}{l}\right)\right],
\end{equation}
where $I_\nu,\,K_\nu$ are the modified Bessel functions of the first and
second kind, respectively, $C_1,\,C_2$ are the integration constants.

It follows from $P(\infty)=\bar P$ boundary condition, that $C_1=0$. The
$C_2$ constant can be found from the spin current continuity condition at
the boundary between the probe and the film
\begin{eqnarray}\label{17}
  &J(R)\equiv\displaystyle\frac{\hbar}{2e}\left\{j(R)P(R)-j_Dl\frac{dP}{dr}\Bigl|_{r=R}\right\}=\nonumber \\
  &=\displaystyle\frac{\hbar}{2e}Q_1j(R)\left(\hat\mathbf M_1\cdot\hat\mathbf M(R)\right),
\end{eqnarray}
where $Q=\displaystyle\frac{\sigma_+-\sigma_-}{\sigma_++\sigma_-}$ is the
the probe spin polarization ($\sigma_\pm$ being the partial
conductivities in the probe), $\hat\mathbf M(R)$ is the unit vector along the film
magnetization near the probe, $\hat\mathbf M_1$ is the same quantity for
the probe. To obtain inverse population of the spin subbands in the film,
$\left(\hat\mathbf M_1\cdot\hat\mathbf M(R)\right)<0$ condition is
necessary.

The substitution of Eq.~(\ref{16}) into Eq.~(\ref{17}) gives
\begin{equation}\label{18}
  C_2=\left\{Q_1 \left(\hat\mathbf M_1\cdot\hat\mathbf M(R)\right)
  -\bar P\right\}\displaystyle\frac{j(R)}{j_D}\left(\frac{R}{l}\right)^\nu
  \frac{1}{K_{\nu+1}\left(\displaystyle\frac{R}{l}\right)},
\end{equation}
so that the electron spin polarization in the film near the probe takes
the form
\begin{equation}\label{19}
  P(r)=\bar P+\left\{Q_1 \left(\hat\mathbf M_1\cdot\hat\mathbf M(R)\right)
  -\bar P\right\}\frac{j(R)}{j_D}\left(\frac{r}{R}\right)^\nu
  \displaystyle\frac{K_\nu\left(\displaystyle\frac{r}{l}\right)}{K_{\nu+1}
  \left(\displaystyle\frac{R}{l}\right)}.
\end{equation}

\begin{figure}
\includegraphics{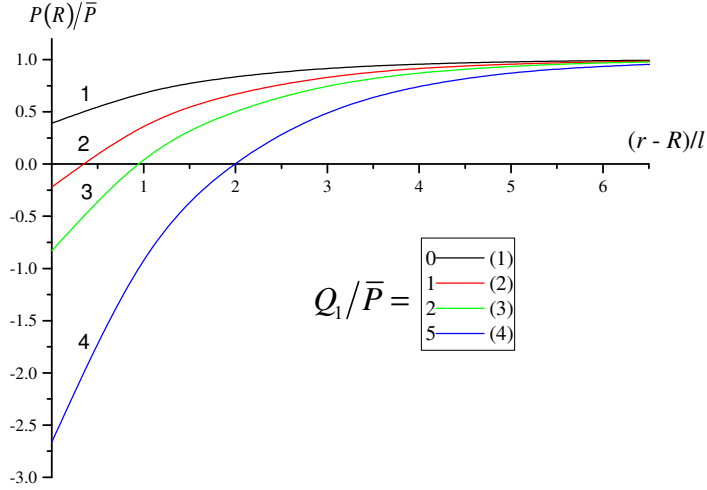}
\caption{Spatial spin polarization distribution near the probe at
$R/l=20,\,j(R)/j_D=1$ and various values of $Q_1/\bar P$ ratio.}\label{fig1}
\end{figure}

It follows from Eq.~(\ref{19}) that the spin polarization approaches
monotonously to the equilibrium value with increasing the distance from
the probe (Fig.~\ref{fig1}). The maximal negative value of the nonequilibrium spin
polarization $\Delta P$ is reached at the probe boundary,
\begin{equation}\label{20}
  \Delta P(R)=\left\{Q_1 \left(\hat\mathbf M_1\cdot\hat\mathbf M(R)\right)
  -\bar P\right\}\displaystyle\frac{j(R)}{j_D}\frac{K_\nu\left(\displaystyle\frac{R}{l}\right)}
  {K_{\nu+1}\left(\displaystyle\frac{R}{l}\right)}.
\end{equation}

As to the dependence of the nonequilibrium spin polarization on the
current density, it is necessary to have in mind that the current density
$j$ appears in Eqs.~(\ref{19}) and~(\ref{20}) not only as an explicit
factor, but also in $\nu$ parameter (see definition~(\ref{15})). Because
of such a reason, the spin polarization near the probe tends to a limiting
value, $P(R)\to Q_1 \left(\hat\mathbf M_1\cdot\hat\mathbf M(R)\right)$.
Note, that the nonequilibrium spin polarization appears in the case of a
nonmagnetic probe also ($Q_1=0$), that means replacing the
spin-polarized electrons with non-polarized ones in the vicinity of the
probe. In that case, the spin polarization remains positive and tends to
zero at high current density ($\Delta P\to-\bar P,\,P\to0$).

In Fig.~\ref{fig1}, a coordinate dependence is shown of the spin
polarization (referred to its equilibrium value) at given values of
$j(R)/j_D$ and $R/l$ ratios and various values of $Q_1/\bar P$ ratio. The
spin polarization near the probe becomes negative at high enough values of
the latter ratio.

\begin{figure}
\includegraphics{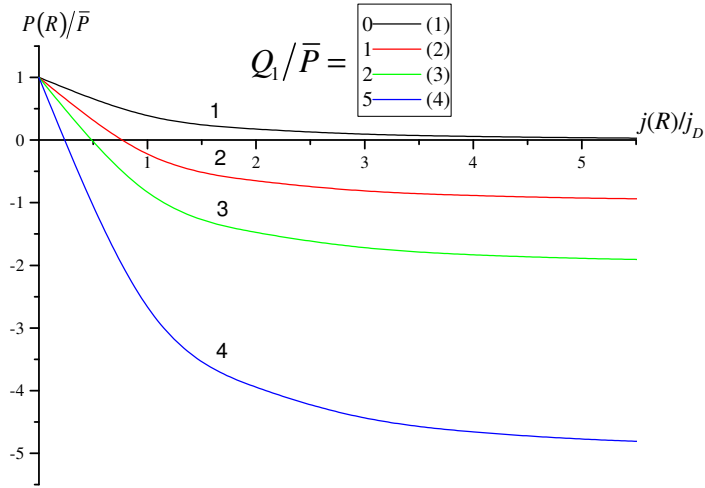}
\caption{Spin polarization at the boundary between the probe and the film
as a function of the (dimensionless) current density $j(R)/j_D$ at
$R/l=20$ and various values of $Q_1/\bar P$ ratio.
}\label{fig2}
\end{figure}

In Fig.~\ref{fig2}, the spin polarization on the probe boundary is shown
as a function of the current density at given value of $R/l$ ratio and various values of $Q_1/\bar P$
ratio.

As numerical analysis shows, the results weakly depend on $R/l$ ratio.
The parameters which influence the spin injection substantially are
$j(R)/j_D$ and $Q_1/\bar P$ ratios.

At $n\sim10^{22}$ cm$^{-3}$, $l\sim10^{-6}$ cm, $\tau\sim10^{-12}$ s, we
have $j_D\sim10^9$ A/cm$^2$. With the current density in the
probe $j_0\sim10^8$ A/cm$^2$ (such a value was reached, e.g.,
in~\cite{Demidov}) and $R/2h\ge10$ (this ratio was equal to 5
in~\cite{Demidov}), we find $j(R)\sim10^9\,\mathrm{A/cm}^2\sim j_D$, that
is sufficient, in accordance with Fig.~\ref{fig2}, for negative spin
polarization near the probe at $Q_1/\bar P\ge1$.

\section{Current-induced magnetic field and sd exchange effective field effects}\label{section4}
At high current densities, influence becomes substantial of the current-induced magnetic
field (the Ampere field). For rough estimates, we may use a simple formula
for the magnetic field of a straight current
\begin{equation}\label{21}
  H=\frac{I}{R}
\end{equation}
without factor 2 of current (``the half of a long wire'').
The substitution of $I=2\pi Rhj(R)$ into Eq.~(\ref{21}) gives the magnetic
field at the probe circumference with given current density:
\begin{equation}\label{22}
  H=2\pi hj(R).
\end{equation}
At $h=10$ nm and $j(R)=10^9$ A/cm$^2$ we have $H\approx600$ Oe.

If the film anisotropy field $H_a$ is lower than the latter value, then
the lattice magnetization near the probe is directed along the probe
circumference. To align the probe magnetization opposite to the film
magnetization, the probe anisotropy field $H_{a1}$ is to be larger than
the indicated $H$ value, and the probe is to be magnetized by such a current of
opposite direction that induces $H>H_{a1}$ field at the probe
circumference.

The inverse population is prevented also with the \emph{sd} exchange
interaction of the injected electrons with the magnetic lattice of the
film, which tends to align the electron spins in the film parallel to ones
in the injector.

The \emph{sd} exchange energy is
\begin{equation}\label{23}
  U_{sd}=-\alpha\int\left(\mathbf M(\mathbf r)\cdot\mathbf m(\mathbf r)\right)\,d^3\mathbf r,
\end{equation}
where $\alpha$ is the dimensionless constant of the \emph{sd} exchange
interaction, $\mathbf M=M\hat\mathbf M$ is the lattice magnetization,
\begin{equation}\label{24}
  \mathbf m=\mu_BnP(\mathbf r)\hat\mathbf M
\end{equation}
is the electron magnetization in the film, $\mu_B$ is the Bohr magneton.

Substitution of Eqs.~(\ref{24}) and~(\ref{19}) into Eq.~(\ref{23}) gives the
following expression for the nonequilibrium part of the \emph{sd} exchange
interaction energy:
\begin{eqnarray}\label{25}
  &\Delta U_{sd}=-\alpha\mu_BnM\cdot2\pi h\int\limits_R^\infty\left\{P(r)-\bar P\right\}r\,dr=\nonumber\\
&=-\alpha\mu_BnM\left\{Q_1\left(\bar\mathbf M_1\cdot\bar\mathbf M\right)
-\bar P\right\}\displaystyle\frac{j(R)}{j_D}\cdot2\pi Rhl.
\end{eqnarray}
The corresponding \emph{sd} exchange effective field
\begin{equation}\label{26}
  \mathbf H_{sd}=-\frac{\delta\Delta U_{sd}}{\delta\mathbf M}
\end{equation}
has an order of magnitude
\begin{equation}\label{27}
  H_{sd}\sim\mu_B\alpha nQ_1\frac{j(R)}{j_D}.
\end{equation}
This field becomes comparable with the molecular field $\sim10^6$ Oe at
$j(R)\sim j_D$. To prevent switching antiparallel initial configuration
($\left(\hat\mathbf M_1\cdot\hat\mathbf M(R)\right)<0$) to parallel one,
it is necessary to pin the magnetization of the film by means of induced
anisotropy with the aid of an antiferromagnetic substrate.

\section{Conclusion}\label{section5}
The analysis within a simplified model shows general possibility of reaching
nonequilibrium negative spin polarization (the spin subband inversion)
with using probe/film structures. However, realization of this possibility
needs some action to prevent switching the antiparallel configuration to
parallel one.

\section*{Acknowledgment}
The work was supported by the Russian Foundation for Basic Research, Grant
No.~08-07-00290.


\begin{thebibliography}{1}
\bibitem{Aronov}
A.~G. Aronov, G.~E. Pikus, Sov. Phys. Semicond. \textbf{10}, 698 (1976).
\bibitem{Slonczewski}
J.~C. Slonczewski, J. Magn. Magn. Mater. \textbf{159}, L1 (1996).
\bibitem{Berger}
L.~Berger, Phys. Rev. B \textbf{54}, 9353 (1996).
\bibitem{Kadigrobov1}
A.~M. Kadigrobov, Z. Ivanov, T. Claeson, R.~I. Shekhter, M. Jonson, Europhys. Lett. \textbf{67},
948 (2004).
\bibitem{Kadigrobov2}
A.~M. Kadigrobov, R.~I. Shekhter, M. Jonson, Low Temp. Phys. \textbf{31}, 352 (2005).
\bibitem{Osipov}
V.~V. Osipov, N.~A. Viglin, J. Commun. Technol. Electron. \textbf{48}, 548 (2003).
\bibitem{Viglin}
N.~A. Viglin, V.~V. Ustinov, V.~V. Osipov, JETP Lett. \textbf{86}, 193 (2007).
\bibitem{Gulyaev1}
Yu.~V. Gulyaev, P.~E. Zilberman, A.~I. Krikunov, A.~I. Panas, E.~M.
Epshtein, JETP Lett. \textbf{85}, 160 (2007).
\bibitem{Gulyaev2}
Yu.~V. Gulyaev, P.~E. Zilberman, E.~M. Epshtein, A.~I. Panas, A.~I. Krikunov,
Patent RU No. 2344528 C1, 20 Jan., 2009.
\bibitem{Chigarev}
S.~G. Chigarev, A.~I. Krikunov, P.~E. Zilberman, A.~I. Panas, E.~M.
Epshtein, J. Commun. Technol. Electron. \textbf{54}, 708 (2009).
\bibitem{Kamke}
E. Kamke, Differentialgleichungen. L\"osungsmethoden und L\"osungen. I.
Gew\"ohnliche Differentialgleichungen. Leipzig, 1959.
\bibitem{Demidov}
V.~E. Demidov, S.~O. Demokritov, G. Reiss, K. Rott, Appl. Phys. Lett. \textbf{90}, 172508
(2007).
\end{thebibliography}
\end{document}